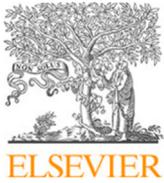
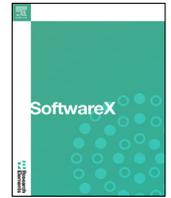

Original software publication

# Nancy: An efficient parallel Network Calculus library

Raffaele Zippo [a,b,*], Giovanni Stea [a]

[a] *Dipartimento di Ingegneria dell'Informazione, University of Pisa, Italy*
[b] *Dipartimento di Ingegneria dell'Informazione, University of Florence, Italy*

**A B S T R A C T**

This paper describes Nancy, a Network Calculus (NC) library that allows users to perform complex min-plus and max-plus algebra operations efficiently. To the best of our knowledge, Nancy is the only open-source library that implements operations working on arbitrary piecewise affine functions, as well as to implement some of them (e.g. sub-additive closure and function composition). Nancy allows researchers to compute NC results using a straightforward syntax, which matches the algebraic one. Moreover, it is designed having computational efficiency in mind: it exploits optimizations of data structures, it uses inheritance to allow for faster algorithms when they are available (e.g., for specific subclasses of functions), and it is natively parallel, thus reaping the benefit of multicore hardware. This makes it usable to solve NC problems which were previously considered beyond the realm of tractable.

## Code metadata

| | |
|---|---|
| Current code version | 1.0.5 |
| Permanent link to code/repository used for this code version | https://github.com/ElsevierSoftwareX/SOFTX-D-22-00130 |
| Permanent link to Reproducible Capsule | |
| Legal Code License | MIT |
| Code versioning system used | git |
| Software code languages, tools, and services used | C# 10, .NET 6.0 |
| Compilation requirements, operating environments & dependencies | .NET 6.0; Linux, MacOS, Windows |
| If available Link to developer documentation/manual | https://nancy.unipi.it/, https://github.com/rzippo/nancy |
| Support email for questions | giovanni.stea@unipi.it, raffaele.zippo@phd.unipi.it |

## Software metadata

| | |
|---|---|
| Current software version | 1.0.5 |
| Permanent link to executables of this version | https://www.nuget.org/packages/Unipi.Nancy/1.0.5 |
| Permanent link to Reproducible Capsule | |
| Legal Software License | MIT |
| Computing platforms/Operating Systems | Linux, MacOS, Windows |
| Installation requirements & dependencies | .NET 6.0 |
| If available, link to user manual - if formally published include a reference to the publication in the reference list | https://nancy.unipi.it/, https://github.com/rzippo/nancy |
| Support email for questions | giovanni.stea@unipi.it, raffaele.zippo@phd.unipi.it |

## 1. Motivation and significance

Worst-case analysis of networked systems is gaining importance due to the emergence of safety-critical applications with real-time requirements, such as factory automation within the Industry 4.0 paradigm, automated or tele-operated driving, coordinated unmanned aerial vehicles, etc. With all these distributed applications, ex-ante certification that the end-to-end network traversal time is always below a known maximum is required to guarantee safety for humans and property. Network Calculus [1] is a well-known theory that uses min-plus and max-plus algebra to infer deterministic worst-case bounds on the delay and backlog of network traffic. It represents traffic as a function of time, and network elements (e.g., regulators, schedulers, links) as operations that modify said functions. Given bounds on the input of a traffic flow (e.g., as enforced by a traffic shaper at the entrance of a network) and knowledge of minimum service given to that flow at network elements (e.g., as enforced by possibly different per-flow schedulers at every hop), NC allows one to compute the maximum delay that traffic from that flow will undergo, under any possible scenario.

* Corresponding author.
 *E-mail address:* raffaele.zippo@phd.unipi.it (Raffaele Zippo).

Algebraically speaking, NC relies on few basic operations, such as minimum, min-plus convolution and sub-additive closure, which can be composed in arbitrary sequences or nested into one another. However, in less-than-trivial cases, pen-and-paper computation of NC expression is not viable, and automated computation of the relevant performance measures (e.g., a maximum delay for a flow traversing a multi-hop network) is instead required. Defining efficient computational representations for functions of time and algorithms implementing NC operations thereupon is by no means a simple task. Works [2,3] discuss the above issue at length, identifying Ultimately Pseudo-Periodic (UPP) functions as the most general class closed with respect to NC operations. These are piecewise affine functions, that have an initial "transient" part, followed by a "period" which is repeated an infinite amount of times. UPP curves arise spontaneously in several practical cases: work [4] shows that they do whenever packetization and finite transmission speeds are factored in. Moreover, flow-controlled networks [5], currently being envisaged for data centers, have UPP service curves, and so do wormhole-routing networks [6] used in systems architectures.

Some NC operations, such as min-plus convolution, have a superquadratic complexity with respect to the number of linear pieces of the operands. Others, such as the sub-additive closure, are exponential. This holds for both the number of elementary operations involved and the number of linear pieces of the result, and implies that chaining a number of NC operations may yield expressions that – while visually and algebraically neat – are algorithmically infeasible, e.g., take days to compute, or may exhaust the hardware resources and not compute at all [5]. This constrains – sometimes severely – the practical uses of NC analysis, and calls for efficient software libraries that implement NC operations.

To the best of our knowledge, there are no public, open-source *libraries* that implement NC operations working on UPP curves. By "library" we mean a set of implementations of min- and max-plus algebra operations, providing ease of constructing and manipulating curves, and allowing a user to specialize algorithms for improved efficiency by adding new code paths or subclasses. The two existing libraries that can handle UPP curves are the RTC Toolbox [7] and the RTaW-Pegase library [8]. The former is a publicly available Java library, whose source code is not available. Thus, one cannot improve on its bugs (e.g. it sometimes cycles infinitely without any discernible cause [9]) or lack of features, such as the lack of sub-additive closure or of a function to find the intersection between two curves (which is essential to speed up several algorithms). RTaW-Pegase is proprietary, and its license does not allow it to be used for benchmarking purposes. An online interpreter is freely available to try its functionalities [10], but its use is limited by the browser interface and license. To the best of our knowledge, the COINC library [11] is no longer available.

There are, instead, several NC *network analysis tools*, i.e., software packages that implement methods of analysis of particular networks. These often implement their own versions of min/max-plus algebra operations, often restricted to classes of curves matching the type of application network they target. A review of their capabilities is reported in [12]. For instance, DEBORAH [13] analyzes FIFO tandems of rate-latency curves traversed by leaky-bucket-shaped flows. As such, it only implements min-plus convolution and subtraction of pseudo-affine curves, both of which are algorithmically trivial. NC-TANDEM-TIGHT [14] analyzes FIFO tandems by modeling their worst-case delay computation as a linear programming problem (hence implicitly ruling out generic UPP curves, which would yield non-convex programs instead). The DiscoDNC tool [15], now NCorg DNC, is limited to ultimately concave/convex piecewise affine curves, for which NC operations are considerably simpler. Work [16] describes a tool written in the NVIDIA CUDA language that computes convolutions and deconvolutions using GPUs, claiming improved efficiency over standard CPU-based computation. However, there seems to be no executable or code to go with this paper. Unfortunately, none of these tools provide reusable, general-purpose NC operations.

In this work, we fill the above gap by releasing Nancy, an open-source NC library, released under the MIT licence, that works with arbitrary UPP curves. Nancy is coded in C#, consists of more than 32k LoC, and is natively parallel. This means that it can exploit multiple cores when available. Most of the NC operations on UPP curves entail comparing affine segments and/or taking minima or maxima, and these operations are easily parallelizable. This allows it to reap a considerable speedup over a single-thread implementation. Furthermore, it implements more efficient algorithms for NC operations working on particular sub-classes of UPP functions (e.g., sub-additive ones), as well as for minimizing the representation of functions, which concur to abate the computation times by orders of magnitude, as shown in [5]. Nancy can be used to implement existing methods of analysis (e.g., Total Flow Analysis or Single Flow Analysis, [17]) used in other tools, with arbitrary UPP curves, i.e. without any of the above-mentioned restrictive hypotheses on the type of curves. Moreover, it can be exploited to devise new methods of analysis, possibly for types of networks that were so far considered computationally untractable, e.g., networks with flow control [5]. Last, but not least, it is designed so that a NC researcher will find it very easy to translate pen-and-paper computations to code: the names used for objects are those which are widespread in the research community, and the syntax is very intuitive.

The Nancy library is expected to be useful to researchers in both academia and industry that face worst-case analysis problems, such as those working in avionics, embedded systems, computer networks at large, and to help system engineers in performance-oriented design. In the rest of the paper, we describe the software design and functionalities in Section 2, we provide examples of how it works in Section 3, and we discuss the impact of the above library in Section 4. Finally, Section 5 concludes the paper.

## 2. Software description

The rest of the paper assumes that the reader is familiar with NC. Those who are not may find a concise introduction to it in [18]. In NC, a *cumulative function* represents the amount of data observed at a network element up to time $t$. A *curve* instead – e.g. a service curve or an arrival curve – represents a property that has to hold in any time interval of length $t$. From a mathematical standpoint, both cumulative functions and NC curves are functions of time. Accordingly, we will often refer to their abscissa as "time" and to their ordinate as "data". The Nancy library provides rich classes that allow one to model, manipulate and run computations with piecewise affine UPP functions. The UPP property states that:

$$f(t + k \cdot d) = f(t) + k \cdot c \ \forall t \geq T, \forall k \in \mathbb{N} \tag{1}$$

The above implies that, in order to compute $f(t)$ for any $t \geq 0$ we only need to store its values in $[0, T + d[$ and the three parameters $T$, $d$ and $c$. This is visualized in Fig. 1. The left plot depicts a function whose transient part (up to $T$) consists of two segments, and whose period (repeated twice) consists of two segments too.

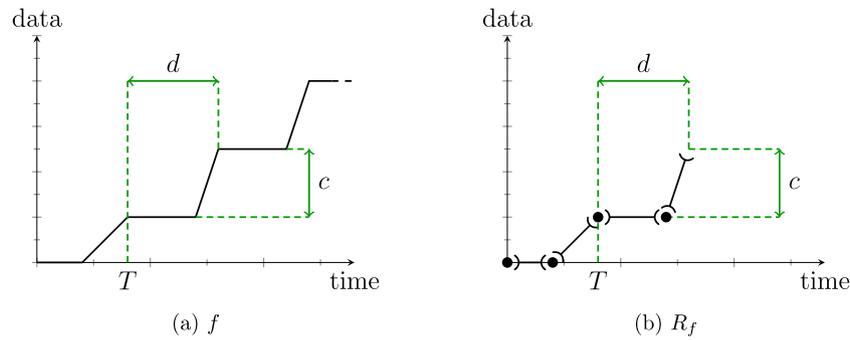

**Fig. 1.** Example of ultimately pseudo-periodic piecewise affine function $f$ and its representation $R_f$, taken from [5].

We represent this in code with the class `Curve`, which is composed of a `BaseSequence` and the values `PseudoPeriodStart` ($T$), `PseudoPeriodLength` ($d$) and `PseudoPeriodHeight` ($c$).

The `BaseSequence` is of type `Sequence`, which models a piecewise affine function in a limited domain, usually left-closed and right-open $[a, b[$ as in [2]. A `Sequence` is composed of `Elements`. An `Element` is an abstract class extended by `Point` (a single time-value pair) and `Segment` (defined in an open interval $]a, b[$). This allows us to model non-continuous UPP curves as well. The right plot of Fig. 1 shows the transient part consisting of four elements, and a period consisting of four elements. Quadruple $\{R_f, T, d, c\}$ is the information required to represent the above function in code.

The numeric type used throughout the library is `Rational`, which represents numbers in $\mathbb{Q} \cup \{+\infty, -\infty\}$. This allows us to perform computations without any approximations (unlike what happens when using the `double` type). To avoid overflow issues, we use `BigInteger`, although we support using `long` instead for better performance, using a compile flag.

These types are *immutable*, i.e. they cannot be modified after construction, and methods will instead create new objects to return. This has a twofold advantage: on one hand, safety of use — an object will never change properties unless it is reassigned explicitly. On the other hand, it favors parallelization, since all computations are independent and can be safely executed concurrently.

Many min- and max-plus operations are implemented as methods of these classes, and are closed [2,3], i.e. they return objects of the same class they belong to — in fact, they compute the bounded UPP representation of the result. This implies that they can be chained, allowing for computation of complex NC expressions. The main operations implemented are:

- `Minimum` and `Maximum`;
- `Addition` and `Subtraction`. The latter can compute both $A(t) - B(t)$ and $[A(t) - B(t)]^+ = max\{A(t) - B(t), 0\}$ (default);
- `Convolution` and `Deconvolution` for the min-plus operations with the same name[1];
- `Vertical-` and `HorizontalDeviation`;
- `MaxPlusConvolution` and `MaxPlusDeconvolution`;
- `Lower-` and `UpperPseudoInverse`;
- `Sub-` and `SuperAdditiveClosure`;

- `Composition`, i.e. $h(t) = f(g(t))$.[2]

We also implement many utility methods and properties for common manipulations and checks, such as:

- Sampling a curve at or "near" a time with `ValueAt`, `LeftLimitAt` and `RightLimitAt`;
- Properties, such as `IsContinuous`, `IsNonDecreasing`, `IsConcave`, `IsSubAdditive`, etc. ;
- Shift along the time axis (`DelayBy`, `AnticipateBy`) and data axis (`VerticalShift`);
- Restriction of a `Curve` to a finite time interval, with the `Cut` operation.

As discussed in [5], the representation of the result of the above operations is often (unavoidably) much larger than necessary — i.e., it includes many more segments than required. This may lead to severe memory/speed inefficiency, especially with chained operations (recall that NC operations often involve algorithms that are superlinear in the number of segments). The Nancy library implements an efficient post-hoc representation minimization algorithm, run by default in all operands before returning the result, which solves this issue. Our minimization algorithm examines the representation of an UPP function and finds the *minimum* number of elements that can represent it exactly, thus minimizing both the transient and the period.[3]

Moreover, we exploit the high parallelizability of the above operations using PLINQ. The Nancy library will parallelize the following:

- aggregate operations, such as the aggregate convolution of a large set of curves;
- large sets of independent tasks. For instance, convolution involves computation of the lower envelope of a set of ppElements over a time domain. Elements are first grouped in (time) buckets according to their interval of definition. Then, per-bucket lower-envelopes are computed independently in parallel.

Lastly, many methods of the `Curve` class are defined as `virtual`, so that its subclasses can provide more efficient algorithms. For instance, the convolution of `ConcaveCurves` is replaced with their minimum, which is less complex. This makes the library easily extensible to exploit existing – or yet-to-be-found – mathematical properties. As an example, three specialized algorithms for `SubAdditiveCurves` have been implemented, which are considerably more efficient: [5, Th. 1], for instance, checks if $f(0) = g(0) = 0$, $f(t) \leq g(t) \forall t$, and $f$ is sub-additive, in which case $f \otimes g = f$. Faster convolution algorithms are also possible when $f$ and $g$ have different asymptotic rates, or when they have the same asymptotic rate and intersect an infinite number of times.

---

[1] Since min-plus operations are more frequent than max-plus ones in NC, "min-plus" is often omitted in the literature when describing these operations. We follow this practice in our software.

[2] This operation is used, for instance, in [19]. To the best of our knowledge, it has been first formally described in the context of UPP curves in [20].

[3] Unlike Finitary RTC [21,22] or the approach in [23], minimization does not require any hypotheses or remove information from the function.

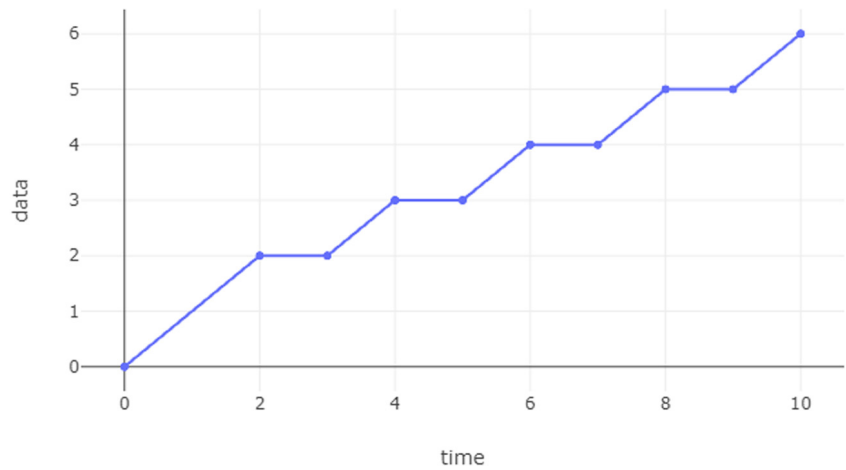

**Fig. 2.** Plot of the curve resulting from Listing 1.

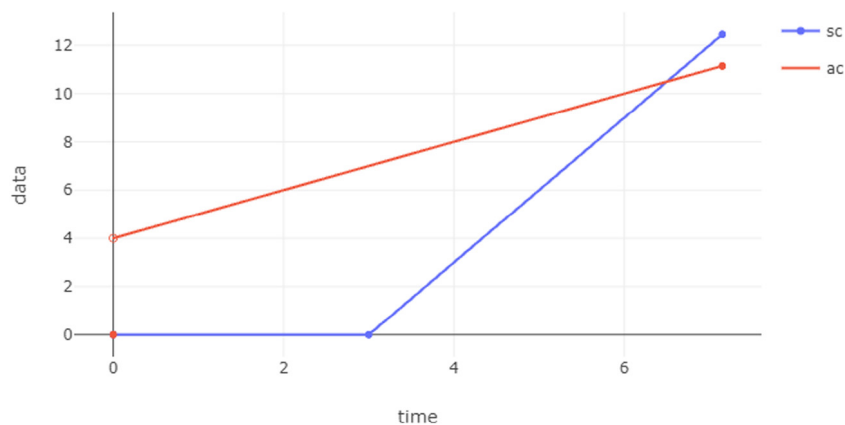

**Fig. 3.** Plot of the curves resulting from Listing 2.

As an example of the above, we report in Table 1 the time it takes to compute the convolution between two sub-additive curves, using both the standard algorithm and the optimized one [5, Th. 1], in both a sequential and a parallel setting. We run the above computations on a laptop computer (i7-10750H, 32 GB RAM). We can clearly see that the algorithm deriving from [5, Th. 1] reduces the computation time by orders of magnitude. The same paper shows that, without using the above-described properties, exact studies of tandems of three or more flow-controlled nodes are just infeasible. This highlights that the flexibility of the type system implemented in Nancy is a great benefit for future improvements. On the other hand, we observe that parallelization brings considerable benefits even for the unoptimized computation. This can be generalized to cases where optimized mathematical properties are not available or not enough (many example of which can be found in [5]). However, in this example, trying to parallelize the optimized algorithm does not yield any benefits: this is to be expected, as the computation time is already below 1 ms, thus the overhead required by parallelization over-compensates any possible time savings. Our library allows users to decide, among other things, whether or not to use parallelization. This is done by modifying the `settings` parameter, and can be used to avoid inefficiencies such as the above one.

**Table 1**
Performance comparison of $\overline{\beta_{416,835} + 313} \otimes \overline{\beta_{552,571} + 970}$, using parallelization and optimized algorithms from [5]. Results are reported for each case as 25th percentile (Q1), median (Q2) and 75th percentile (Q3) over 10 independent runs.

|  | Standard | Optimized |
|---|---|---|
| Sequential | Q1: 20512.67 ms<br>Q2: 20569.76 ms<br>Q3: 20733.97 ms | Q1: 0.16 ms<br>Q2: 0.17 ms<br>Q3: 0.64 ms |
| Parallelized | Q1: 5860.67 ms<br>Q2: 6030.53 ms<br>Q3: 7657.40 ms | Q1: 0.49 ms<br>Q2: 0.52 ms<br>Q3: 2.22 ms |

## 3. Illustrative examples

In this section we show how the Nancy library can be used to easily build curves and compute NC expressions. These selected examples are included in the GitHub repository as .NET Interactive Notebooks, where we use `XPlot.Plotly` to visualize the computed curves, as shown in the figures below. Our aim here is to show that the library syntax, which is rather intuitive, closely matches NC algebra, making it very easy for a practitioner to compute (and possibly plot) the results of its expressions.

The first example in Listing 1 and Fig. 2 shows how to construct a generic `Curve`, providing the limited `Sequence` description and the pseudo-period parameters. The library also provides simplified constructors for common types of curves, as shown in Listing 2 and Fig. 3.

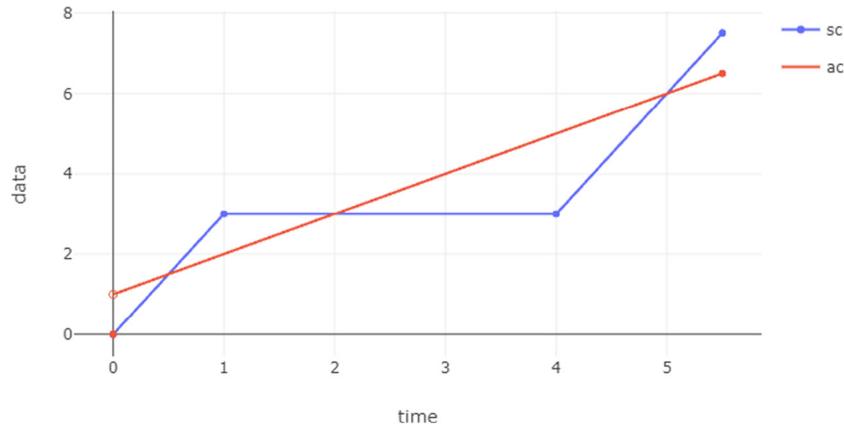

**Fig. 4.** Plot of the two curves in Listing 3. The maximum horizontal deviation is equal to $d((2, 2), (4, 2)) = 2$.

```
var c = new Curve(
    baseSequence: new Sequence(new Element[]
    {
        new Point(time: 0, value: 0),
        new Segment(startTime: 0, endTime: 2, rightLimitAtStartTime:0, slope: 1),
        new Point(2, 2),
        new Segment(2, 3, 2, 0),
        new Point(3, 2),
        new Segment(3, 4, 2, 1)
    }),
    pseudoPeriodStart: 2,
    pseudoPeriodLength: 2,
    pseudoPeriodHeight: 1
);
```

**Listing 1:** Constructing a Curve using the generic constructor.

```
var sc = new RateLatencyServiceCurve(rate: 3, latency: 3);
var ac = new SigmaRhoArrivalCurve(sigma: 4, rho: 1);
```

**Listing 2:** Code example using simplified constructors.

```
var sc = Curve.Minimum(
    new RateLatencyServiceCurve(3, 0),
    new RateLatencyServiceCurve(3, 4) + 3
);
var ac = new SigmaRhoArrivalCurve(1, 1);
var delay = Curve.HorizontalDeviation(ac, sc);
Console.WriteLine(delay); // Output: 2/1
```

**Listing 3:** Constructing non-trivial curves and computing the delay bound as the maximum horizontal deviation.

```
var beta = new RateLatencyServiceCurve(3, 2);
var alpha = new SigmaRhoArrivalCurve(3, 2);
var theta = 4;
var delta_theta = new DelayServiceCurve(theta);

var alpha_theta = Curve.Convolution(alpha, delta_theta);
var diff = Curve.Subtraction(beta, alpha_theta, nonNegative: true);
var residual = Curve.Minimum(diff, delta_theta);
```

**Listing 4:** Computing a residual service curve for a flow in a FIFO aggregate.

We can then use such curves to compute more complex curve expressions, or delay bounds, as shown in the example of Listing 3 and Fig. 4 (from [3, p. 121]).

We now show two examples of more elaborate expressions taken from the NC literature. In Listing 4 and Fig. 5 we show the computation of a residual service curve for a flow in a FIFO aggregate, whose expression is ([1, Ch. 6]):

$$\beta^\theta = [\beta - \alpha \otimes \delta_\theta]^+ \wedge \delta_\theta,$$

where $\beta$ is the aggregate service curve, $\alpha$ is the arrival curve of the cross-flow, $\delta_x$ is a right-shift by $x$ and $\theta \geq 0$ is a time parameter.

Our last example is based on the recent work in [19, Th. 1], which computes a strict service curve for Interleaved Weighted Round Robin (IWRR) through multiple steps. In Listing 5 and Fig. 6, we implement this computation using the parameters in [19, Fig. 3]. Our documentation website includes other examples, including a complete study that shows how to compute the delay bound of a leaky-bucket-constrained flow traversing a tandem of IWRR nodes. Note that this requires convolution of non-convex service curves, like the one in Fig. 6.

## 4. Impact

Despite being 30 years old, Network Calculus is still a flowering field of research. A search on Scopus reports 758 scientific papers published in the last 10 years (80 in 2021 only) where "Network Calculus" appears in the title, abstract or keywords. Research groups working on it exist in both the academia (at least in Europe, US, Canada, China, India, Brazil) and the industry (components, avionics, systems). It is in particular industrial interest that has sparked a renewed attention in NC in the last few years, with the advent of new standards (e.g., Time-Sensitive Networks) and new scenarios and opportunities (e.g., the possibility of running real-time control algorithms of physical objects from the Cloud or

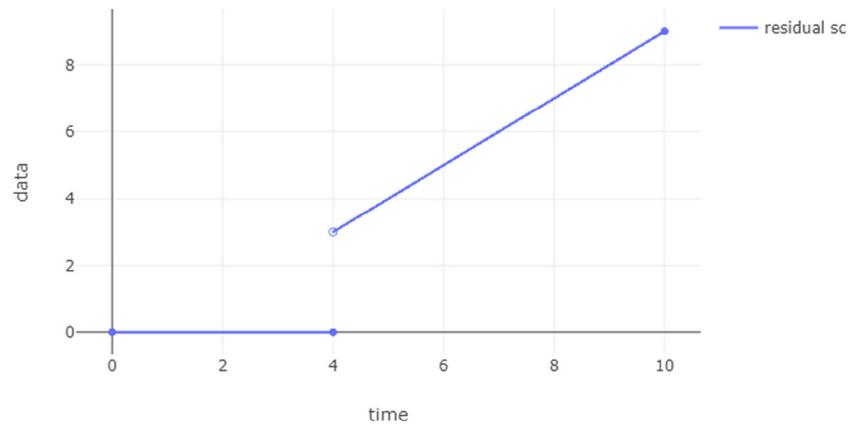

**Fig. 5.** Plot of the residual service curve resulting from Listing 4.

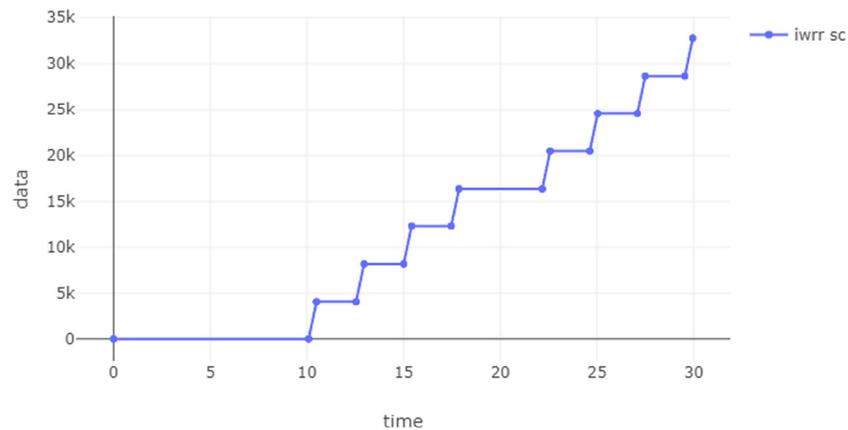

**Fig. 6.** Plot of the service curve resulting from Listing 5.

```
var weights = new []{4, 6, 7, 10};
var l_min = new []{4096, 3072, 4608, 3072};
var l_max = new []{8704, 5632, 6656, 8192};
var beta = new RateLatencyServiceCurve(
    rate: 10000, // 10 Mb/s, but using ms as time unit
    latency: 0
);
var unit_rate = new RateLatencyServiceCurve(1, 0);

int Phi_i_j(int i, int j, int x) {...}

int Psi_i(int i, int x) {...}

int L_tot(int i) {...}

int i = 0; // the flow of interest
var stairs = new List<Curve>();
for(int k = 0; k < weights[i]; k++)
{
    var stair = new StairCurve(l_min[i], L_tot(i));
    var delayed_stair = stair.DelayBy(Psi_i(i, k * l_min[i]));
    stairs.Add( delayed_stair );
}
var U_i = Curve.Addition(stairs); // summation of min-plus curves
var gamma_i = Curve.Convolution(unit_rate, U_i);
var beta_i = Curve.Composition(gamma_i, beta);
```

**Listing 5:** Computing IWRR's strict service curve. The code to compute the intermediate parameters is collapsed for the sake of conciseness.

the emergence of Digital Twins). The Nancy library presented in this paper can be used by all the above actors to address their research needs. It has already been used in research projects (namely, a joint project between Arm Ltd and the University of Pisa), as well as to write scientific papers [5].

## 5. Conclusions

This paper has presented Nancy, a C# library for Network Calculus operations, written for NC researchers. One of the features provided by the Nancy library is the attention to computational efficiency. As already discussed, NC is widely believed to be computationally hard, especially when generic UPP curves are concerned. This has severely limited its applicability to practical problems so far. Quite often, researchers have resorted to pessimistic bounding techniques in order to obtain computable results, or have been forced to limit the scope of their analysis to few network nodes. Nancy tackles the problem of computational efficiency in several concurrent ways. The first one is *inheritance*: while algorithms for UPP curves at large are known, research is currently discovering subclasses of UPP curves whose properties allow more efficient algorithms. One such examples are sub-additive curves, discussed in [5], where the cost of convolution can be abated by orders of magnitude. The second one is built-in *optimization* of the curve data structures. It is shown in [5] that NC operations yield results whose UPP representation is redundant. Optimizing that representation ex-post comes at negligible cost, and may save orders of magnitude of computations (besides memory occupancy, which often becomes a problem itself). The third one is native parallel implementation, which harvests all the underlying computational power, distributing both NC operations and sub-tasks thereof. Future work will consider improving the efficiency of NC algorithms, as well as investigating ways to integrate Nancy with existing NC tools, e.g., as described in [24], or to leverage specialized hardware support, e.g., GPUs [16].

**Declaration of competing interest**

The authors declare the following financial interests/personal relationships which may be considered as potential competing interests: Giovanni Stea reports financial support and administrative support were provided by ARM Ltd. Giovanni Stea reports a relationship with ARM Ltd. that includes: funding grants.

**Data availability**

No data was used for the research described in the article.

**Acknowledgments**

This work was supported in part by the Italian Ministry of Education and Research (MIUR) in the framework of the CrossLab project (Departments of Excellence), and by the University of Pisa, Italy, through grant "Analisi di reti complesse: dalla teoria alle applicazioni" - PRA 2020. We are grateful to ARM Ltd., and in particular to Dr. Matteo Andreozzi, for their help in publishing this software.